\documentstyle[aps,prl,twocolumn,epsf]{revtex}
\tighten

\begin{document}
\draft

\title{Equilibration and freeze-out in an exploding system}
\author{J. P. Bondorf$^1$, H. Feldmeier$^2$, I. N. Mishustin$^{1,3,4}$, G. Neergaard$^{1,5}$}
\address{$^1$The Niels Bohr Institute, Blegdamsvej 17, DK-2100 Copenhagen, Denmark.\\
$^2$GSI, Planckstrasse 1, D-64291 Darmstadt, Germany.\\
$^3$Institute for Theoretical Physics, Goethe University, Frankfurt am Main, D-60054, Germany.\\
$^4$The Kurchatov Institute, Russian Research Center, 123182 Moscow, Russia.\\
$^5$Institute of Physics and Astronomy, University of Aarhus, Denmark.}
\maketitle

{\it Abstract.}
We use a simple gas model to study non-equilibrium aspects
of the multiparticle dynamics relevant to heavy ion collisions.
By performing numerical simulations for various initial conditions
we identify several characteristic features of the fast dynamics occurring
in implosion-explosion like processes.
\vspace{0.5cm}

{\it Introduction.}
Thermodynamic equilibrium at an intermediate stage of a heavy ion collision is 
frequently a basic assumption in models of the colliding nuclear matter.
These models range from statistical models of nuclear multifragmentation 
to the fluid dynamical models of the quark gluon plasma. 
In contrast, microscopic models
of molecular dynamics type (e.g.\ RQMD, FMD and NMD), which are based 
upon constituent interactions do not have this assumption built in. 
Such models are appropriate for testing to what extent 
thermodynamic equilibrium is actually achieved.

Previously we have used the Nuclear Molecular Dynamics (NMD) 
model~\cite{selforg} 
to describe the collision dynamics and clusterization process 
in nucleonic matter. In a special study of the 
time evolution of a finite system of colliding particles we noticed that 
the phase space distance $d$ between particles in the 
combined coordinate and momentum space has a distinct time evolution
for different groups of particles:
For nucleons which form clusters, 
$d$ is initially large and rapidly decreasing until it stays
small and rather constant from a certain time. This can be interpreted 
as freeze-out of the hard core scattering interaction. This 
interpretation was supported by the fact that for interactions with no
hard core scattering the phase space distance for clusters 
always remained small.
Averaged over all nucleons the phase space distance $d$ grows with time
reflecting the overall expansion of the system.

The aim of the present paper is to study the behavior of a finite
piece of exploding matter before and around freeze-out 
of hard core interaction.
Before and at this stage, the evolution of the matter is highly dynamical, 
leading to characteristic space-time distributions of freeze-out and flow.
Whether the dynamics actually results in thermodynamical equilibrium 
before freeze-out is not clear a priori. 
Different aspects of this problem have been studied previously
within various approaches~\cite{randrup,baym,henning,bondorf}.
In a previous paper~\cite{bolproc} we employed a new variable, the
pseudo entropy, to study the equilibration.
In the present letter we use theoretically well-founded concepts 
to investigate the dynamics of the exploding matter.

{\it Classical  gas  model.}
To this end we study the dynamical properties of a simple
gas model, where the constituent particles interact like billard balls
by classical Newtonian dynamics. 
A first application of this model of nuclear collisions was made 
in~\cite{firstapp}.
We consider a gas of identical classical balls of radius $r_c$.
They perform classical non-relativistic hard-sphere 
elastic scatterings at the impact parameter $b \leq 2r_c$
with conservation of energy, momentum and angular momentum.
(The balls do not have intrinsic rotation.)
The initial configuration consists of 
$A$ such particles placed randomly within a
sphere of radius $R$, rejecting configurations where particles 
overlap within the hard core distance. 
This puts a constraint on $R$ which should not be too small.  

We prepare the system at certain energy and
density, and let it evolve until the interactions 
between the constituents cease.
We further impose spherical symmetry on the system.
In this way we minimize specific complications of the interactions 
and the geometry,
and we can concentrate on a few basic properties of the behavior of small 
shortlived systems, which we believe are of general interest for heavy 
ion physics.

We study the evolution of the matter in the gas ball 
in a simple dynamical process, 
which is initiated by ingoing flow of the particles (implosion).
In this situation the system will first compress and
then expand, thus simulating some basic features of a heavy ion collision.

The initial velocities of the particles are chosen as a
superposition of thermal and collective motion.
We use a spherically symmetric linear profile for the initial collective
velocity field
\begin{equation}
\vec{v}(\vec{r}) = v_{0f}\,\vec{r}/R
\label{hubbleflow}
\end{equation}
where $v_{0f}$ is a model parameter.
In our simulations we fix the total energy $E=E_{fl}+E_{th}$,
and vary the fraction $\eta$ of the flow energy, $\eta=E_{fl}/E$,
where $E_{fl}$ and $E_{th}$ are the flow energy and the thermal energy,
respectively.
Because of the way in which the system is built up, these energies
will fluctuate from event to event with a relative uncertainty 
of order $A^{-\frac{1}{2}}$.   

We use conventional nuclear scales: 
the mass of the constituent particle is $940$~MeV,
the hard core radius is $r_c=0.4\,-\,0.5$~fm, 
and the initial radius of the gas sphere is $R=1.2\,A^{1/3}$~fm.
We have performed simulations with $A=50$ and $A=100$ varying
the flow velocity parameter $v_{0f}$ from $-0.5$ to $0.5$
(this could be in units of the velocity of light, $c=1$, 
but the velocity scale is in fact arbitrary in this model), and 
varying $\eta$ from $0$ (thermal explosion) to $1$ (implosion).

\begin{figure}
\epsfxsize=8.6truecm
\epsfbox{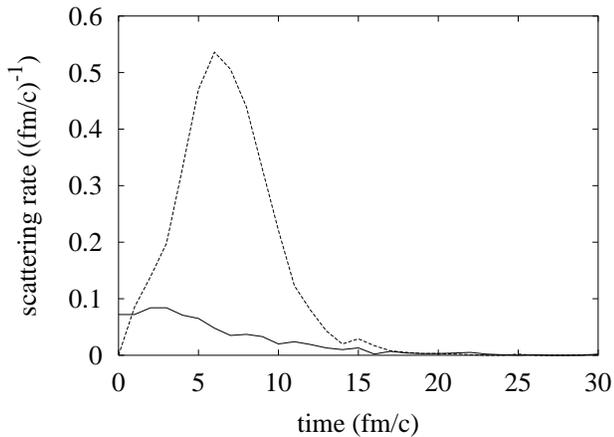}
\caption[]{The total scattering rate for a thermal explosion (solid curve)
and an implosion (dashed curve) with $v_{0f}=-0.5$.
Here, and in Figs.~\ref{fig:frout025} and \ref{fig:frf} $r_c=0.5$~fm.}
\label{fig:sr}
\end{figure}

{\it Freeze-out.}
A very important characteristics of evolving matter
is the scattering rate $\nu(t)$, defined as
the total number of scatterings per particle per unit time.
It is obvious that in an implosion $\nu(t)$
first grows and then decreases, while in a thermal explosion it decreases. 
In Fig.~\ref{fig:sr} shows $\nu(t)$ for the two cases.
The scattering rate roughly follows the matter density, because
the mean free path $\lambda$ is inversely proportional to the density $n$ 
($\lambda=1/n\sigma$, $\sigma$ being the scattering cross section). 

One of the most fundamental questions in heavy ion physics is
{\it when can a collection of particles be considered as coherent matter?}
In our simplified model we shall consider particles which collide with
other particles as the matter, whereas particles which 
do not scatter any more are decoupled from the matter, 
or {\it frozen out}. 
Therefore we study the space-time distribution of the last scattering points
(the {\it freeze-out field}), because this defines the space-time
limit of the coherent matter.

\begin{figure}
\epsfxsize=8.6truecm
\epsfbox{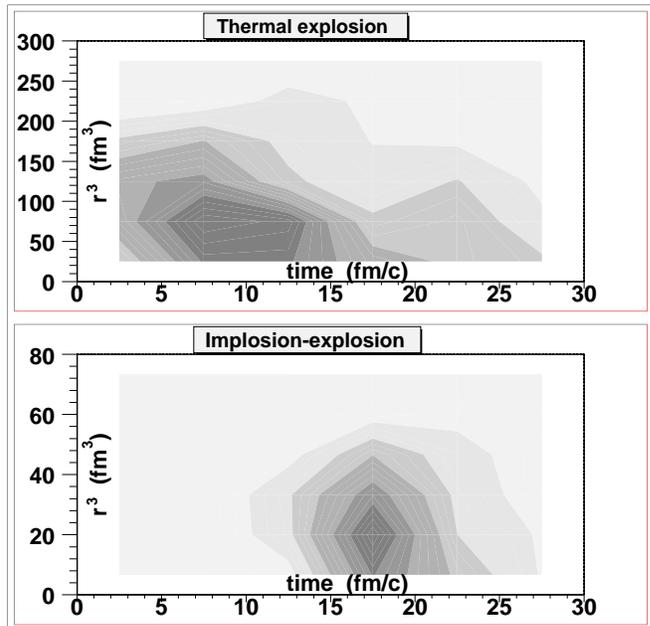}
\caption[]{
Freeze-out fields for a thermal explosion ($\eta=0$, top panel),
and for an implosion ($\eta=1$, lower panel), with 
$100$ particles. $v_{0f}=-0.25$. 
We show the distribution in the $(t,r^3)$ plane so that a uniform
distribution of points in the plot corresponds to a uniform 
distribution in the 3 dimensional space.
The distributions are based on summing the results of 10 events.
In the thermal explosion, a total of $362$ last scatterings took place
in the $10$ events, whereas in the implosion all particles scattered 
at least once in each event ($500$ scatterings in total).}
\label{fig:frout025}
\end{figure}

Two examples of a freeze-out field are shown in Fig.~\ref{fig:frout025}. 
It is seen that the distribution is rather spread, 
in particular for the thermal explosion.
This disagrees with the Cooper-Frye picture~\cite{cooperfrye}, 
which assumes a sharp freeze-out hypersurface. 
A similar conclusion was reached in~\cite{bravina}
for microscopic simulations of relativistic heavy ion collisions.

An attractive way of understanding the freeze-out process is to think
of the field of most recent scatterings (the {\it scattering field}): 
At any given time, there will be points 
in space where each particle had its most recent scattering.
The distribution of these points in space-time is the scattering field 
$l(\vec{r},t;t')$, where
$\vec{r}$ and $t$ are space-time coordinates 
and $t'$ is the present time (the instant at which we look at the system).
The entire field $l$ as function of $\vec{r}$ and $t$ will of course change 
with time $t'$, but only until complete freeze-out when
the scattering field becomes fully fixed 
and equal to $l(\vec{r},t;\infty)$.

\begin{figure}
\epsfxsize=8.6truecm
\epsfbox{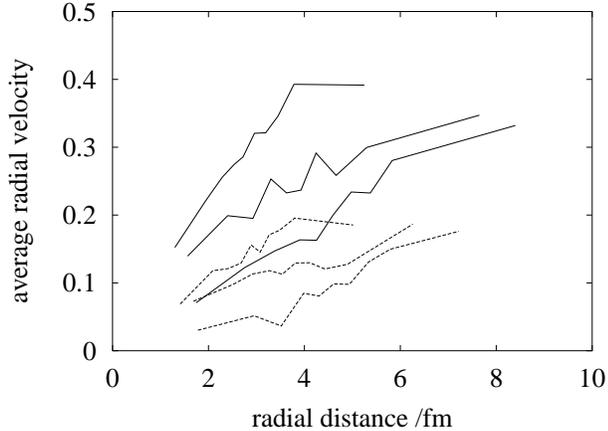}
\caption[]{
Time integrated freeze-out flow fields 
for $v_{0f}=-0.25$ (the three dashed curves) and
$v_{0f}=-0.5$ (the three solid curves). 
Within each set of curves the flow fraction $\eta$ in the 
initial state is varied, $\eta=0,0.5,1$ (bottom to top).}
\label{fig:frf}
\end{figure}

In addition to the freeze-out field itself, we can consider other variables
defined on the freeze-out space-time coordinates.
Of particular importance is the {\it freeze-out flow field}, 
since this may be related to the experimentally measured flow.
The freeze-out flow field should be defined as a non-thermal part 
of the particle velocity
at the freeze-out points. In general the definition of collective flow
is somewhat ambiguous. However, in our case of spherical geometry it
can be defined as the mean radial component of the 
particle velocities at the freeze-out points.
In Fig.~\ref{fig:frf} we show some examples of time integrated freeze-out
flow fields. Any non-zero value of these curves is a sign of collective flow.
The general picture is that the freeze-out flow increases with the ingoing flow
fraction $\eta$ and with the distance to the center of the explosion.
Note also that even for the thermal explosion ($\eta=0$), 
the freeze-out flow is non-zero.
One sees from the figure that within statistical errors the two
families of curves reveal the scaling with the initial velocity.

{\it Equilibration.}
We do not wish to make any assumptions about 
equilibrium, be it global or local, partial or complete.
We will, however, make use of the concept of entropy, 
since this can formally be calculated also out of equilibrium.

To study one-body observables, we
reduce the $6A$ dimensional phase-space of the $A$ particles 
to the one-body phase-space in $6$ dimensions 
in the standard way~\cite{kennard}.
We further exploit the radial symmetry of the system by reducing
the phase-space to two dimensions ($|\vec{r}|$,$|\vec{p}|$), the 
advantage being a gain in statistics.
Then we introduce a finite grid in this reduced phase-space,
dividing each of the two axes into a number of segments.
But instead of working with a fixed grid in phase-space, 
we let the entire grid expand or contract with the majority of particles.
Our experience shows that  
$R_{20}=(1/A\,\sum_{n=1}^{A} r_n^{20})^{1/20}$
is a good measure of the spatial extension of the system.
This means that the physical size $\Omega_i$ 
(in units of $(2\pi\hbar)^3$)
of cell $i$ in phase-space varies with time.
We choose our grid such that all cells have equal phase space volume.

One can use the phase-space distribution of particles obtained from 
microscopic simulations to characterize the degree of equilibrium.
First of all, for given initial conditions we perform several
independent dynamical runs, and accumulate the points in different
phase-space cells at each time step.
In this way we obtain the occupation probability distribution
$\{p_i\}$, where
\begin{equation}
p_i=\frac{{\mathrm number\ of\ points\ in\ box}\ i}{\mathrm total\ number\ of\ points\ in\ phase\ space}.
\label{pi}
\end{equation}
Then we repeat the simulations for a reference system where particles have
sufficient time for equilibration in a fixed volume $4\pi/3R_{20}^3$,
where $R_{20}$ is calculated from the actual spatial distribution
of particles in the dynamical simulations. In this way we obtain
the corresponding equilibrium distribution $\{p_i^{eq}\}$.
We then construct the quantity
\begin{equation}
S = - \sum_i p_i \log(p_i/\Omega_i),
\label{S}
\end{equation}
which in the case of equilibrium would be the real entropy~\cite{ll}.
Dividing by $\Omega_i$ inside the logarithm
eliminates the trivial trend that a larger phase space cell 
holds more points than a smaller cell. 
As a measure of equilibrium we consider the quantity
\begin{equation}
\Sigma = \exp(S-S_{ref}).
\label{eqmeasure}
\end{equation}
$\Sigma$ is thus a measure of how close the actual
phase-space distribution is to that
of an equilibrized system with the same volume and energy.
$\Sigma=1$ cooresponds to full equilibrium.

Since the grid expands or contracts with the particles, 
the information on flow is somewhat erased. 
One would thus expect $\Sigma$ to reach a constant value after freeze-out.

\begin{figure}
\epsfxsize=8.6truecm
\epsfbox{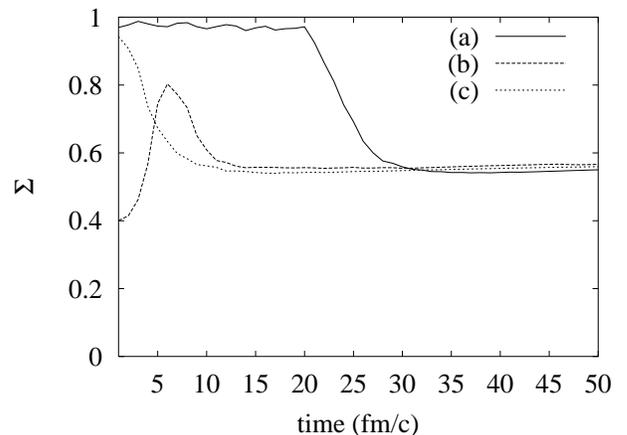}
\caption[]{The equilibration measure $\Sigma$ calculated for 
the three cases discussed in the text.
The results are based on an emsemble of $20$ events with $50$ particles.
Here, $r_c=0.4$~fm.}
\label{fig:sigma}
\end{figure}

In Fig.~\ref{fig:sigma} we show the behavior of the equilibrium measure
$\Sigma$ in the following three cases: 
(a): The particles are kept inside a spherical container 
of radius $R=4.4$~fm
until $t=20$~fm/c, when the container walls are removed;
(b): $100$\% ingoing flow ($\eta=1$), 
intended to simulate an implosion-explosion process;
(c): The particles are started with $50$\% thermal 
energy and $50$\% outgoing flow ($\eta=0.5$), simulating an explosion
from a not fully thermalized state.
We notice the following features of $\Sigma$ in this figure: 
In the case (a) $\Sigma \simeq 1$ as long as particles are
kept in the container, reflecting the fact that the system is 
in equilibrium. Then when the container is removed, the system goes
out of equilibrium and $\Sigma$ decreases until collisions between
the particles cease. This is as intuitively expected. 
Now, in the implosion-explosion case (b), we do not know in 
advance if the system will reach a state of equilibrium or not.
Here, $\Sigma$ starts from a low value, 
since the initial state is far from equilibrium.
As collisions tend to equilibrize the system, $\Sigma$ increases, but
does not reach the equilibrium $\Sigma=1$.
However, additional analyses show that the particles obtain 
a nearly Maxwellian velocity distribution 
from $t=6$~fm/c, which is also the time of maximum compression.
During the de-equilibration time, $\Sigma$ decreases,
and the rate of decrease reflects the dynamics of freeze-out.
It is interesting to note that the thermal explosion (a) 
and the ``non-thermal explosion'' (c) both approach 
the same asymptotic value $\Sigma \simeq 0.55$,
as does the implosion-explosion case.
Therefore we cannot use the asymtotic behavior of $\Sigma$ to conclude 
about the degree of equilibrium at early stages of the system evolution.
The asymptotic value of $\Sigma$ is not universal, for instance for
$100$\% outgoing flow $\Sigma$ is approximately constant $\simeq 0.4$.

{\it Conclusions.}
We have performed dynamical simulations within a simple
gas model and analysed the results both with regard to freeze-out,
flow, and degree of equilibrium.
The space-time distributions of last scatterings are broad in both 
space and time.  Without using any assumptions about equilibrium
we have defined the freeze-out collective flow, 
at least for the simple geometry adopted here.
A general trend is that the freeze-out distribution is sharper
and the freeze-out flow is stronger
when the system goes through a state of high compression.
Our calculations show that the maximum compression coincides
with the maximum degree of equilibrium, but even at this stage
equilibration is not complete.
The present investigation uses a very schematic interaction.
In the future we are planning to extend the study to other types
of interaction.

Our general conclusion is that small systems evolving under fast
dynamics exhibit many features which cannot be described adequately by
equilibrium concepts.

\vspace{0.5cm}

{\it Acknowledgments.}
This work was in part supported by the Danish Natural Science Research Council.
JPB thanks the Nuclear Theory Group at GSI Darmstadt, 
where part of this work was done.
GN thanks the Niels Bohr Institute, and
and The Leon Rosenfeld Foundation and the L{\o}rup Foundations
for financial support.
INM thanks The Niels Bohr Institute, and 
The Humbold Foundation for financial support.


\end{document}